\documentstyle[twoside,fleqn,espcrc2,epsf]{article}
\def\b{\beta}

\def\g{\gamma}

\def\p{\pi}       
\def\r{\rho}      
\def\s{\sigma}    

\def\D{\Delta}

\newcommand{\ra}{\rightarrow}

\newcommand{\be}{\begin{equation}}
\newcommand{\ee}{\end{equation}}
\newcommand{\bea}{\begin{eqnarray}}
\newcommand{\eea}{\end{eqnarray}}

\newcommand{\beq}{\begin{equation}}
\newcommand{\eeq}{\end{equation}}
\newcommand{\cc}{\cite}
\newcommand{\lb}{\label}

\def \3{\ss}

\newcommand{\AmS}{{\protect\the\textfont2
  A\kern-.1667em\lower.5ex\hbox{M}\kern-.125emS}}

\title{Quenched Hadron Spectroscopy with Improved Actions}

\author{Wolfgang Bock\address{Center for Computational Physics,   
        University of Tsukuba,  
        305 Ibaraki, Japan}%
}
       
\begin{document}

\begin{abstract}
A variety of different combinations of 
improved gluon  and fermion actions is tested 
for quenched hadron spectroscopy. 
\end{abstract}

\maketitle
\section{INTRODUCTION}
The improvement of the lattice QCD action has become recently 
a main focus of research in lattice field theory. 
Improved actions for lattice QCD are designed to remove lattice-spacing 
artifacts in numerical estimates of physical observables. This allows 
to conduct the numerical calculations on much coarser and smaller lattices 
and thereby to substantially reduce the simulation costs.                                        
In this contribution a variety of different combinations of gauge 
and fermion actions is tested for quenched hadron spectroscopy. For the 
gluon action we used the tadpole improved L\"uscher-Weisz 
(T-LW) \cc{LuWe85,LeMa93,AlDi95} and the Iwasaki (IW) action \cc{Iw83}. 
The LW action has been obtained by extending Symanzik's 
perturbative improvement program to lattice gauge theories, demanding $O(a^2)$ improvement of 
on-mass-shell quantities \cc{LuWe85}. The use of tad-pole improvement 
is essential to enhance the convergence of the perturbative expansion and extend 
perturbation theory to larger distances \cc{LeMa93}.  
The IW action on the contrary is a renormalization group improved action. 
Scaling tests carried out recently for both models have 
been very encouraging \cc{AlDi95,KaIw96,BoIw96}. 
%
%
The  fermion actions which we shall consider in this contribution are the 
Sheikholeslami-Wohlert (SW) action \cc{ShWo85}, the 
D234 action \cc{AlKl95} and for comparison also the standard Wilson action (W). 
Symanzik improvement of the Wilson action leads to the SW action which eliminates 
the cut-off effects in on-mass-shell quantities to order $O(a)$.
With the D234 action one attempts to go even one step beyond the $O(a)$ improvement and to 
eliminate partially some of the cut-off effects to $O(a^2)$. 
Also in the case of the improved fermion actions it is essential to couple the 
fermions to a "tadpole improved" (T)
lattice link variable \cc{LeMa93}. 
We computed the hadron spectrum for the following five combinations 
of improved gluon and fermion actions: 
T-LW\&W, T-LW\&T-SW, T-LW\&T-D234, IW\&T-SW and IW\&T-D234.
\section{DETAILS OF THE SIMULATION}
In order to be able to compare the results obtained with the different gluon actions 
we have conducted the simulations for the T-LW and IW actions at 
the critical $\b$ of the deconfining phase transition for 
$N_t=2$ \cc{BoIw96}. We performed the simulations on a $6^3 16$ lattice using
a standard Hybrid Monte Carlo algorithm. 
For the determination of the hadron spectrum we generated about 500 
configurations, which were separated by 50 trajectories. 
For the inversion of the 
fermion matrix we used the BICG$\g_5$ algorithm which differs from
the CG algorithm by the insertion of a $\g_5$ in the inner products.
We find that the BICG$\g_5$ is, within the interval $m_{\pi}=0.7$-$1.0$,
by a factor $3$-$4$ times faster  
than the BICGstab and by a factor $6$-$8$ faster than the CG algorithm. 
The BICG$\g_5$ algorithm  
is not guaranteed to converge as a division by zero 
can occur during the iteration process. We have however never 
encountered a breakdown of the algorithm although we performed in total 
about $2.5\;10^5$ inversions of the fermion matrix.
%
\begin{figure} [ttt]
\vspace{-0.5cm}
\centerline{ \epsfysize=7.5cm \epsfbox{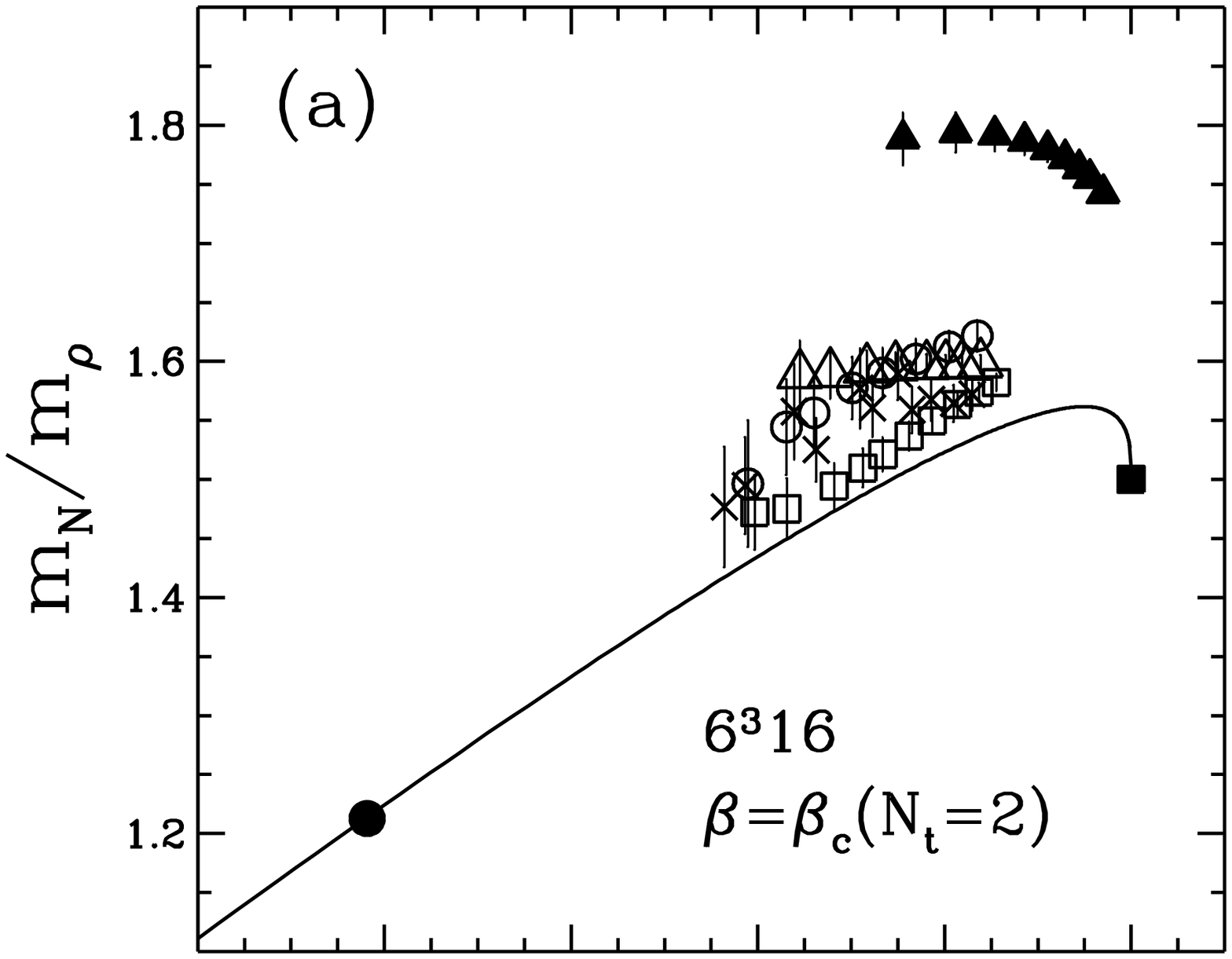} }
\vspace{-2.2cm}
\centerline{ \epsfysize=7.5cm \epsfbox{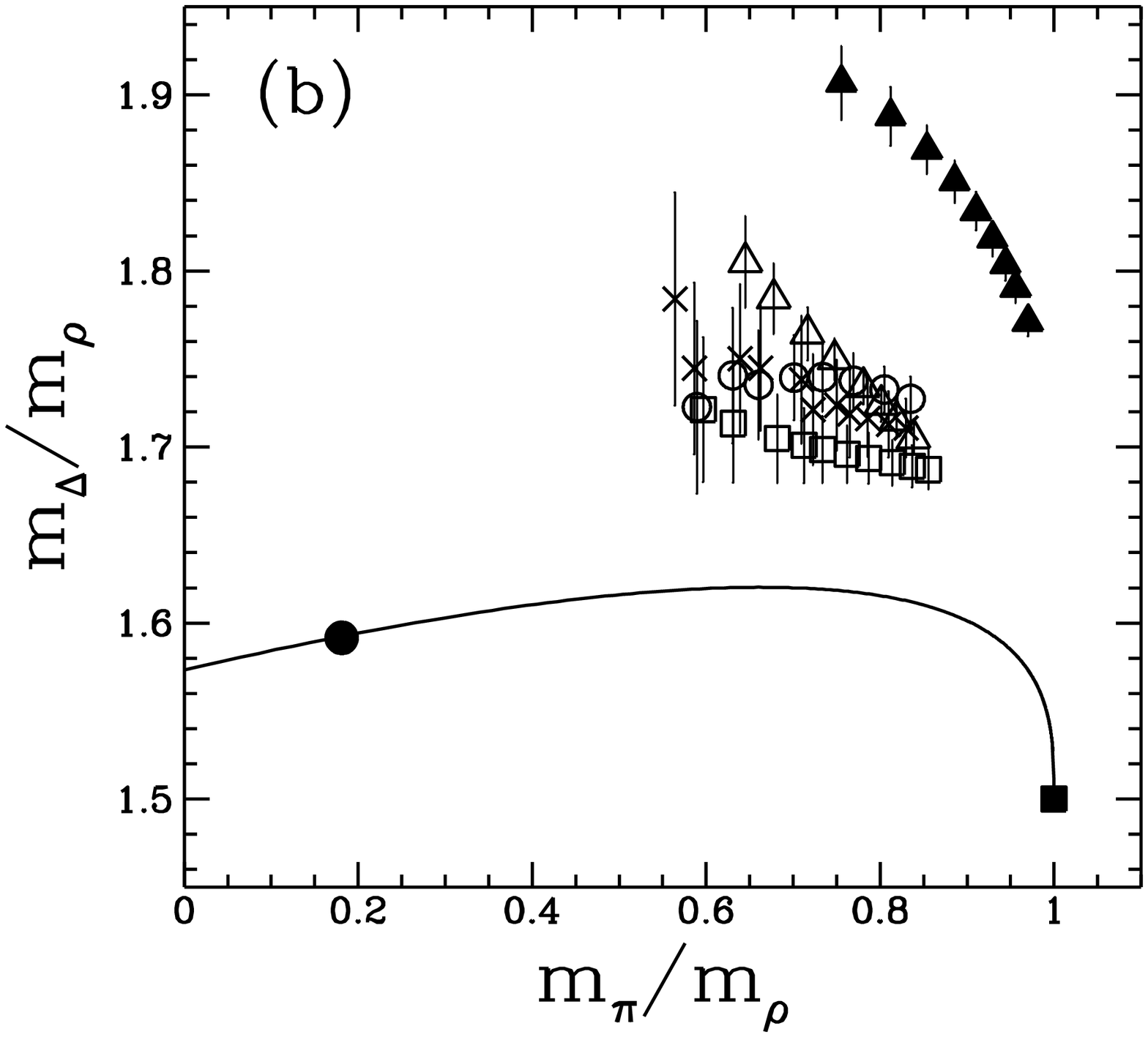} }
\vspace{-1.8cm}
\caption{The mass ratios $m_N/m_{\r}$ (a) and $m_{\D}/m_{\r}$ (b) 
as a function of $m_{\p}/m_{\r}$.     
The T-LW{\rm\&}W, T-LW{\rm\&}T-SW, T-LW{\rm\&}T-D234, IW{\rm\&}T-SW and IW{\rm\&}T-D234 data
are marked by the filled triangles, open circles, crosses, open triangles and
squares. The horizontal error bars are dropped as they are smaller than the symbol size.}
\lb{FIG1}
\vspace{-0.5cm}
\end{figure}

For the determination of the hadron spectrum we 
implemented the standard correlation functions 
for the $\pi$, $\r$, $a_0$, $a_1$ and $b_1$ mesons, the nucleon (N) and the $\D$,                  
using local operators for the quark field, both at the source and the sink.            
The meson and baryon propagators have been fitted with   
$A_M [\exp(-m_M t) + \exp(-m_M (T-t))]$ and $A_B \exp(-m_B t)$. 
We varied $t_{{\rm min}}$ systematically and averaged the masses 
when it was difficult to pick out the best among equally good fits. 
The mass spectrum has been determined in the interval $m_{\pi}\!=\!0.7$-$1.5$, where 
finite size effects are very small. All correlation functions gave 
a very clear and stable signal in this pion mass range,   
expect for the $a_0$, $a_1$ and $b_1$ channels. We discovered however 
that the signal in those channels improves significantly at larger pion masses, and   
therefore extended the calculation of the $a_0$, $a_1$ and $b_1$ correlation functions with          
the IW\&T-D234 action, and for comparison also with the IW\&W action,
to somewhat larger pion masses, $m_{\p}\!=\!1.5$-$2.0$.
Apart from the hadron mass spectrum we also computed the string tension $\s$ which 
we determined from the exponential fall-off of the Polyakov loop 
correlation function.
\begin{figure} [ttt]
\vspace{-0.5cm}
%
%
\centerline{ \epsfysize=7.5cm \epsfbox{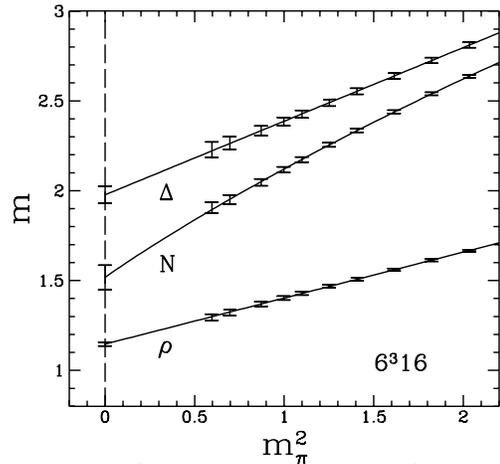} }
\vspace{-2.0cm}
\caption{ The masses $m_\r$, $m_N$ and $m_\D$ as a function of $m_\p^2$ for the
IW{\rm \&}D234 action.}
\lb{FIG2}
\vspace{-0.8cm}
\end{figure}
\section{RESULTS}
In fig.~1 we have plotted the mass ratios $m_N/m_\r$ (a) and $m_\D/m_\r$ (b) as 
a function of $m_\p/m_\r$. The filled circles to the left 
correspond to the physical point and the filled  squares to the 
right to the heavy quark limit, where $m_{\pi}/m_{\rho}=1$
and $m_{N}/m_{\rho}\!=\!m_{\D}/m_{\rho}\!=\!3/2$.                                          
The solid lines connecting the two points represent the result of a   
phenomenological quark model formula derived in ref.~\cc{On78}.          
The graphs show that the points 
which correspond to the combinations T-LW\&T-SW (open circle), T-LW\&T-D234 (crosses), 
IW\&T-SW (open triangles) and IW\&T-D234 (squares)
cluster in a narrow strip which is located in both cases above the solid line. 
The large gap between the standard Wilson action data (filled triangles) and all other 
data reveals very clearly the dramatic effect of the fermion improvement.                            
A naive linear extrapolation of the T-LW\&T-SW, T-LW\&T-D234 and IW\&T-D234
data in fig.~1a down to the physical value of $m_{\pi}/m_{\rho}$ 
leads to a $m_N/m_\r$ value which is not too far off from the experimental point (filled circle). 
The extrapolation of the IW\&T-SW data on the other hand results in a value that is significantly larger
indicating that the cut-off effects are bigger for this particular action. 
We checked that our data obtained with the improved fermion actions are,  
within error bars, consistent with the standard Wilson action results which  
\begin{figure} [ttt]
\vspace{-0.5cm}
\centerline{ \epsfysize=7.5cm \epsfbox{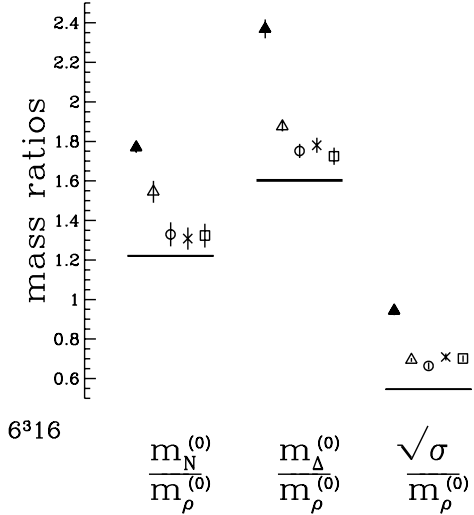} }
\vspace{-1.0cm}
\caption{Mass ratios for the various actions.                       
The symbols have the same meaning as in fig.~1.}
\lb{FIG3}
\vspace{-0.6cm}
\end{figure}
have been obtained during the past years on a much finer and bigger lattices.

Using chiral perturbation theory 
we extrapolated the hadron masses $m_\r$, $m_{a_0}$, $m_{a_1}$, $m_{b_1}$,  
$m_N$ and $m_\D$ to the chiral limit, $m_{\p} \ra0$.                                                  
To this extend we fitted the meson mass data $m_M$ 
with $m_M^{(0)}+ c_2 m_\p^2$, where 
$m_M^{(0)}$ designates the mass in the chiral limit 
and $M\!=\!\r,a_0,a_1,b_1$ in our case. For the nucleon it is essential to include 
also a term in the chiral fit which is cubic in the pion mass,
$m_N^{(0)}+ c_2 m_\p^2 +c_3 m_\p^3$. Our results show that the coefficient $c_3$            
depends strongly on the particular action. It is largest for the three combinations 
T-LW\&T-SW, T-LW\&T-D234 and IW\&T-D234 and roughly by a factor two smaller for 
the other two actions \cc{Bo96}. In the case of the $\D$ we fitted the 
mass data with $m_\D^{(0)}+ c_2 m_\p^2$, as a cubic fit leads to a $c_3$ value 
which is consistent with zero \cc{Bo96}.           
As an example, we plotted in fig.~2 the $m_\r$, $m_N$ and $m_\D$ data for the case of 
the IW\&T-D234 action as a function of $m_\p^2$. 
The results of the fits are represented in this graph by the solid lines. 
 
The ratios of the extrapolated masses, $m_{N}^{(0)}/m_{\r}^{(0)}$ and $m_{\Delta}^{(0)}/m_{\r}^{(0)}$           
are displayed in fig.~3 along  with $\sqrt{\s}/m_{\r}^{(0)}$. 
The horizontal lines represent the experimental values. We find that 
the T-LW\&T-SW (open circle), T-LW\&T-D234 (crosses) and IW\&T-D234 
(squares) mass ratios are very close to the experimental values.
The gap is slightly bigger for the IW\&T-SW data 
(open triangles), and largest for the  
standard fermion action data (filled triangles).
The results for the mass ratios $m_{a_0}^{(0)}/m_{\r}^{(0)}$,
$m_{a_1}^{(0)}/m_{\r}^{(0)}$ and $m_{b_1}^{(0)}/m_{\r}^{(0)}$ are summarized in table~1 which 
also shows that the mass ratios obtained with the improved fermion action
are much closer to the experimental values than the ones obtained with 
the standard fermion action. 
\begin{table}[t]
\vspace{-0.2cm}
\label{TAB1}
\begin{center}
\begin{tabular}{|c|c|c|c|c|}
\hline
Table~1          & IW\&W       & IW\&        & experiment  \\ 
                 &             & T-D234      &            \\ \hline           
$m_{a_1}^{(0)}/m_{\r}^{(0)}$ & $2.37(15)$  & $1.81(8)$   & 1.600(52)   \\ 
$m_{a_0}^{(0)}/m_{\r}^{(0)}$ & $2.37(17)$  & $1.70(10)$  & 1.279(2)    \\ 
$m_{b_1}^{(0)}/m_{\r}^{(0)}$ & $2.28(14)$  & $1.90(10)$  & 1.601(13)   \\ \hline 
\end{tabular}
\end{center}
\vspace{-1.0cm}
\end{table}

I thank P.~de~Forcrand, Y.~Iwasaki, K.~Kanaya, P.~Mackenzie, T.~Yoshie and A.~Ukawa
for many interesting discussions.
The simulations were performed on the VPP500/30 at the University of Tsukuba. 
This research was supported by the Japanese Society for the Promotion of Science. 
\vspace{-0.1cm}

\end{document}